\documentclass[a4paper,10pt]{article}
\usepackage{amssymb}
\usepackage{amsmath}
\usepackage{epsfig}
\usepackage{subfigure}
\usepackage{graphics}

\usepackage{latexsym}
\usepackage{rotating}
\usepackage{titlesec}
\newcommand{\squeezeup}{\vspace{-9.5mm}}

\title{On SK and KK Integrable Systems}

\author{Metin G\"{u}rses \thanks{gurses@fen.bilkent.edu.tr}\\
{\small Department of Mathematics, Faculty of Science}\\
{\small Bilkent University, 06800 Ankara - Turkey}\\
Asl{\i} Pekcan \thanks{Email:aslipekcan@hacettepe.edu.tr} \\
{\small Department of Mathematics, Faculty of Science} \\
{\small Hacettepe University, 06800 Ankara - Turkey}
}

\date{\nonumber}
\begin{document}
\maketitle
\date{\nonumber}
\newtheorem{thm}{Theorem}[section]
\newtheorem{Le}{Lemma}[section]
\newtheorem{defi}{Definition}[section]
\newtheorem{ex}{Example}[section]
\newtheorem{pro}{Proposition}[section]
\baselineskip 17pt

\numberwithin{equation}{section}

\begin{abstract}
To obtain new integrable nonlinear differential equations there are some well-known methods such as Lax equations with different Lax representations. There are also some other methods which are based on integrable scalar nonlinear partial differential equations. We show that some systems of integrable equations published recently are the ${\cal M}_{2}$-extension of integrable scalar equations. For illustration we give Korteweg-de Vries, Kaup-Kupershmidt, and Sawada-Kotera equations as examples. By the use of such an extension of integrable scalar equations we obtain some new integrable systems with recursion operators. We give also the soliton solutions of the system equations and integrable standard nonlocal and shifted nonlocal reductions of these systems.

\end{abstract}


\section{Introduction}
There are many ways of obtaining new integrable systems of equations such as taking the Lax representations in algebras of higher rank. In these methods we obtain real and complex valued coupled nonlinear equations which possess recursion operators and Hirota bilinear forms. There are also some other methods which use integrable systems with less number of dynamical variables to produce integrable systems with more dynamical variables.
 Recently we observe such an effort to obtain systems of Sawada-Kotera (SK) and Kaup-Kuperschmidt (KK) equations \cite{liang0}, \cite{liang1} by the use of Lax representations. The purpose of this paper is to show that the systems obtained in \cite{liang0}, \cite{liang1} are easily obtained by a method that we call ${\cal M}_{2}$-extensions of the SK and KK equations. This method is so general that it can be used to any integrable scalar equation.

Let ${\cal M}_{2}$ be a special subclass of $2 \times 2$ matrices. Let $A \in {\cal M}_{2}$ and be given by $A=a_{1}\,I+a_{2}\, \Sigma$ where $a_{1}$ and $a_{2}$ are independent components of the matrix $A$, $I$  is  the $2 \times 2$ identity matrix, and
\begin{equation}
\Sigma=\left(\begin{array}{ll}
        0 & \sigma \cr
        1 & 0 \cr
        \end{array} \right),
\end{equation}
where $\sigma$ is a real constant. Since $\Sigma^2=\sigma \,I$, products of all such matrices are also in this subclass of matrices. Hence ${\cal M}_{2}$ is a commutative subclass of $2 \times 2$ matrices. Due to the special properties of this subclass ${\cal M}_{2}$ of $2 \times 2$ matrices we can generate new systems of integrable equations from scalar integrable equations. We call this method as ${\cal M}_{2}$-extension of integrable scalar equations. As an illustration consider the well-known Korteweg-de Vries (KdV) equation $u_{t}=u_{xxx}+6 u u_{x}$ with Lax pair $L=D^2+u$, ${\cal A}=4D^3+6uD+3u_x$ so that the Lax equation is satisfied by virtue of the KdV equation
\begin{equation}
L_t+[L,{\cal A}]=0,
\end{equation}
and the recursion operator ${\cal R}_{KdV}=D^2+4u+2u_{x} D^{-1}$ \cite{KdV}. Here $D$ is the total $x$-derivative and $D^{-1}=\int^x$ is the standard
anti-derivative. Then ${\cal M}_{2}$-extension of the KdV equation, its Lax pair, and recursion operator are respectively given as
\begin{eqnarray}
&&U_{t}=U_{xxx}+6 U U_{x}, \label{denk0}\\
&&L=I D^2+U, \\
&&{\cal A}=4ID^3+6UD+3U_x, \\
&&{\mathcal R}=I D^2+4U +2U_{x}\, D^{-1},
\end{eqnarray}
where
\begin{equation}
U=uI+v \Sigma=\left(\begin{array}{ll}
        u & \sigma v \cr
        v & u \cr
        \end{array} \right).
\end{equation}
The pair $L$ and ${\cal A}$ solves the Lax equation due to the ${\cal M}_{2}$-extension of the KdV equation (\ref{denk0}) for $U$. In componentwise the above equations give a system of equations for the dynamical variables $u$ and $v$
\begin{eqnarray}
&&u_{t}=u_{xxx}+6(uu_{x}+\sigma vv_{x}), \\
&&v_{t}=v_{xxx}+6(uv)_{x},
\end{eqnarray}
admitting the recursion operator
\begin{equation}
{\cal R}=\left(\begin{array}{ll}
        R_{KdV} & \sigma (4 v+2 v_{x}\, D^{-1})\cr
        4 v+2 v_{x}\, D^{-1} & R_{KdV} \cr
        \end{array} \right).
\end{equation}
It is interesting that the above extension of the KdV equation is equivalent to the (pseudo)-complexification of the KdV equation. First let $\sigma =\epsilon |\sigma|$ where $\epsilon =0, \pm 1$. Then scale $v \to \frac{v}{\sqrt{|\sigma|}}$ (for $ \sigma \ne 0$) so the above system becomes
\begin{eqnarray}
&&u_{t}=u_{xxx}+6(uu_{x}+\epsilon vv_{x}), \\
&&v_{t}=v_{xxx}+6(uv)_{x}.
\end{eqnarray}

\vspace{0.3cm}
\noindent
{\bf Remark 1.} For $\epsilon=0$ the above system is the extension of the KdV with its linearized equation for $v=\delta u$. For $\epsilon \ne 0$  this system is a consequence of the (pseudo)-complexification $u \to U=u+e v$ of the KdV equation where $e$ is the pseudo-complex unit $e^2=\epsilon$.
For complex numbers $\epsilon=-1$ but for pseudo-complex numbers $\epsilon=1$. Complex conjugation for both cases is the same $e^{*}=-e$. Hence our conclusion is that
${\cal M}_{2}$-extension of the KdV equation is the unification of linearization and  pseudo-complexification of the KdV equation. Our second conclusion is that this is valid in general.

We use ${\cal M}_{2}$-extension method to integrable scalar equations to obtain systems of integrable equations and to obtain new integrable nonlocal  equations. This method  consists of three main steps. The first step is to replace the dynamical variable of the integrable scalar equation by $u I+v \Sigma$ where $u$ and $v$ are the dynamical variables of the system. Here, since $\Sigma^2=\sigma I$ the system of equations contain also the constant $\sigma$.  We write the dynamical equations for $u$ and $v$. By using the recursion operator of the scalar equation we obtain the recursion operator of the system equations for $u$ and $v$. Furthermore, if the Hirota bilinear form of the scalar equation is known then we obtain the bilinear form of the system of equations. At this step we obtain an integrable system of equations for $u$ and $v$. Second step is to obtain the symmetrical version of the system equations by defining new dynamical variables $q=u+v$ and $r=u-v$. At the same time one can obtain the recursion operator with respect to the dynamical variables $q$ and $r$. The third step is to apply consistent reductions; standard (unshifted) nonlocal reductions $r(x,t)=q(\epsilon_{1} x, \epsilon_{2} t)$ and $r(x,t)=\bar{q}(\epsilon_{1} x, \epsilon_{2} t)$, and shifted nonlocal reductions $r(x,t)=q(\epsilon_{1} x+x_0, \epsilon_{2} t+t_0)$ for $x_0, t_0 \in \mathbb{C}$ and $r(x,t)=\bar{q}(\epsilon_{1} x+x_0, \epsilon_{2} t+t_0)$ for $x_0, t_0 \in \mathbb{R}$ where $\epsilon_{1}^2=\epsilon_{2}^2=1$ to obtain standard nonlocal and shifted nonlocal reductions of the system for $q$ and $r$ \cite{abl1}-\cite{WT}. Using the reduction formulas, we can obtain the recursion operators of the nonlocal differential equations. In general these equations will be new and integrable. Soliton solutions of the standard nonlocal and shifted nonlocal equations can be easily obtained by using soliton solutions of the systems and reduction formulas.

In the following section we find ${\cal M}_{2}$ extensions of the SK and KK equations, their recursion operators, and symmetrical versions of these systems. In Section 3 we obtain nonlocal reductions of the SK and KK systems. In Section 4 we find shifted nonlocal reductions of SK and KK sytems. In Section 5 we present Hirota bilinearization of SK and KK systems and give one-soliton solutions of the equations.

\section{SK and KK systems}

Recently, we observe some publications on the extensions of higher order integrable equations by using the Lax representations \cite{liang0}, \cite{liang1}. Here we show that such extensions are nothing but the ${\cal M}_{2}$-extension of an integrable scalar equation explained in the previous section.

Scalar versions of SK and KK equations are given respectively as \cite{liang0}, \cite{liang1}, \cite{gur}-\cite{kaup}:

\vspace{0.3cm}
\noindent
(1) SK equation, Lax pair, and recursion operator
\begin{eqnarray}
&& u_{t}+u_{5x}+5 u u_{xxx}+5 u_{x} u_{xx}+5 u^2 u_{x}=0, \label{SKeqn} \\
&&L=D^3+uD, \\
&&{\cal A}=9 D^5+15 u D^3+15 u_{x} D^2+5(u^2+2 u_{xx}) D,\\
&& {\cal R}_{SK}=D^6+6u D^4+9 u_{x} D^3+(9 u^2+11 u_{xx}) D^2+(10 u_{xxx}+21 u u_{x}) D\nonumber\\
&&\hspace{0.8cm}+4 u^3+16 u u_{xx}+6u_{x}^2+5 u_{4x}+u_{x} D^{-1}\,(2 u_{xx}+u^2)-u_{t}\, D^{-1}.
\end{eqnarray}

\vspace{0.3cm}
\noindent
(2) KK equation, Lax pair, and recursion operator
\begin{eqnarray}
&& u_{t}+u_{5x}+10 u u_{xxx}+25 u_{x} u_{xx}+20 u^2 u_{x}=0, \\
&& L=D^3+2uD+u_{x},
\end{eqnarray}
\begin{eqnarray}
&&{\cal A}=9 D^5+3 u D^3+45 u_{x} D^2+(20u^2+35 u_{xx}) D +10(u^2+u_{xx}),\\
&& {\cal R}_{KK}=D^6+12 u D^4+36 u_{x} D^3+(36 u^2+49 u_{xx}) D^2+(35 u_{xxx}+120 u u_{x}) D\nonumber\\
&&\hspace{0.8cm}+32 u^3+82 u u_{xx}+69 u_{x}^2+13 u_{4x}+2u_{x} D^{-1}\,( u_{xx}+4 u^2)-2u_{t}\, D^{-1}.
\end{eqnarray}

\vspace{0.3cm}
\noindent
${\cal M}_{2}$-extensions of these equations, Lax pairs, and recursion operators are respectively given as

\vspace{0.3cm}
\noindent
(3) SK system, Lax pair, and recursion operator
\begin{eqnarray}
&& U_{t}+U_{5x}+5 U U_{xxx}+5 U_{x} U_{xx}+5 U^2 U_{x}=0, \\
&&L=ID^3+UD, \\
&&{\cal A}=9 ID^5+15 U D^3+15 U_{x} D^2+5(U^2+2 U_{xx}) D,\\
&& {\cal R}=ID^6+6U D^4+9 U_{x} D^3+(9 U^2+11 U_{xx}) D^2+(10 U_{xxx}+21 U U_{x}) D\nonumber\\
&&\hspace{0.8cm}+4 U^3+16 U U_{xx}+6U_{x}^2+5 U_{4x}+U_{x} D^{-1}\,(2 U_{xx}+U^2)-U_{t}\, D^{-1}.
\end{eqnarray}
The above equations correspond to the system of equations for the dynamical variables $u$ and $v$ as
\begin{eqnarray}
&&u_t+u_{5x}+5uu_{xxx}+5u_xu_{xx}+5u^2u_x+5\sigma(vv_{xx}+v^2u)_x=0,\label{SKa}\\
&&v_t+v_{5x}+5(vu_{xx}+uv_{xx}+u^2v)_x+5\sigma v^2v_x=0.\label{SKb}
\end{eqnarray}
The recursion operator of this system is
\begin{equation}\label{SK-rec}
{\cal R}=\left(\begin{array}{ll}
        R_{SK}+\sigma\,A_{11} & \sigma A_{21}\cr
        A_{21} & R_{SK}+\sigma\,A_{11} \cr
        \end{array} \right),
\end{equation}
where
\begin{align}
&A_{11}=9v^2D^2+21vv_xD+12uv^2+16vv_{xx}+6v_x^2+u_xD^{-1}v^2+2v_xD^{-1}(v_{xx}+uv),\\
&A_{21}=-v_t+6vD^4+9v_xD^3+(18uv+11v_{xx})D^2+(10v_{xxx}+21(uv)_x)D+4(\sigma v^3+3u^2v)\nonumber\\
&\hspace{0.8cm}+16(vu_{xx}+uv_{xx})+12u_xv_x+5v_{4x}+v_xD^{-1}(2u_{xx}+u^2+\sigma v^2)+2u_xD^{-1}(v_{xx}+uv).
\end{align}
By letting $u=q+r$, $v=q-r$, and $t\rightarrow at$, where $a$ is a constant, we get symmetrical version of the system (\ref{SKa}) and (\ref{SKb}) as
\begin{align}
&aq_t+q_{5x}+\frac{5}{2}[(\sigma+3)q_{xx}-(\sigma-1)r_{xx}]q_x+10[(\sigma+1)q^2-(\sigma-1)qr]q_x+5(\sigma-1)(r^2-q^2)r_x\nonumber\\
&+\frac{5}{2}(\sigma-1)(r_{xx}-q_{xx})r_x+\frac{5}{2}[(\sigma+3)q-(\sigma-1)r]q_{xxx}+\frac{5}{2}(\sigma-1)(r-q)r_{xxx}=0,\label{symSKa}\\
&ar_t+r_{5x}+\frac{5}{2}[(\sigma+3)r_{xx}-(\sigma-1)q_{xx}]r_x+10[(\sigma+1)r^2-(\sigma-1)qr]r_x+5(\sigma-1)(q^2-r^2)q_x\nonumber\\
&+\frac{5}{2}(\sigma-1)(q_{xx}-r_{xx})q_x+\frac{5}{2}[(\sigma+3)r-(\sigma-1)q]r_{xxx}+\frac{5}{2}(\sigma-1)(q-r)q_{xxx}=0.\label{symSKb}
\end{align}

\vspace{0.3cm}
\noindent
(4) KK system, Lax pair, and recursion operator
\begin{eqnarray}
&& U_{t}+U_{5x}+10 U U_{xxx}+25 U_{x} U_{xx}+20 U^2 U_{x}=0, \\
&&L=ID^3+2UD+U_{x}, \\
&&{\cal A}=9ID^5+3 U D^3+45 U_{x} D^2+(20U^2+35 U_{xx}) D +10(U^2+U_{xx}),\\
&& {\cal R}=ID^6+12 U D^4+36 U_{x} D^3+(36 U^2+49 U_{xx}) D^2+(35 U_{xxx}+120 U U_{x}) D\nonumber\\
&&\hspace{0.8cm}+32 U^3+82 U U_{xx}+69 U_{x}^2+13 U_{4x}+2U_{x} D^{-1}\,( U_{xx}+4 U^2)-2U_{t}\, D^{-1}.
\end{eqnarray}
In componentwise the above equations give the following system of equations for the dynamical variables $u$ and $v$
\begin{align}
&u_t+u_{5x}+10uu_{xxx}+25u_xu_{xx}+20u^2u_x+5\sigma(2vv_{xxx}+5v_xv_{xx}+4(v^2u)_x)=0,\label{KKa}\\
&v_t+v_{5x}+5(2vu_{xxx}+5v_xu_{xx}+2uv_{xxx}+5u_xv_{xx}+4(u^2v)_x)+20\sigma v^2v_x=0.\label{KKb}
\end{align}
The recursion operator of the above system is
\begin{equation}\label{KK-rec}
{\cal R}=\left(\begin{array}{ll}
        R_{KK}+\sigma \,A_{11} & \sigma A_{21}\cr
        A_{21} & R_{KK}+\sigma\,A_{11} \cr
        \end{array} \right),
\end{equation}
where
\begin{align}
&A_{11}=36v^2D^2+120vv_xD+96uv^2+82vv_{xx}+69v_x^2+8u_xD^{-1}v^2+2v_xD^{-1}(v_{xx}+8uv),\\
&A_{21}=-2v_t+12vD^4+36v_xD^3+(72uv+49v_{xx})D^2+(35v_{xxx}+120(uv)_x)D\nonumber\\
&\hspace{0.8cm}+32(\sigma v^3+3u^2v)+82(vu_{xx}+uv_{xx})+138u_xv_x+13v_{4x}+2v_xD^{-1}(u_{xx}+4u^2+\sigma 4v^2)
\nonumber\\
&\hspace{0.8cm}+2u_xD^{-1}(v_{xx}+8uv).
\end{align}
Letting $u=q+r$, $v=q-r$, and $t\rightarrow at$, where $a$ is a constant, yields symmetrical version of the system (\ref{KKa}) and (\ref{KKb}) as
\begin{align}
&aq_t+q_{5x}+\frac{25}{2}[(\sigma+3)q_{xx}-(\sigma-1)r_{xx}]q_x+40[(\sigma+1)q^2-(\sigma-1)qr]q_x+20(\sigma-1)(r^2-q^2)r_x\nonumber\\
&+\frac{25}{2}(\sigma-1)(r_{xx}-q_{xx})r_x+5[(\sigma+3)q-(\sigma-1)r]q_{xxx}+5(\sigma-1)(r-q)r_{xxx}=0,\label{symKKa}\\
&ar_t+r_{5x}+\frac{25}{2}[(\sigma+3)r_{xx}-(\sigma-1)q_{xx}]r_x+40[(\sigma+1)r^2-(\sigma-1)qr]r_x+20(\sigma-1)(q^2-r^2)q_x\nonumber\\
&+\frac{25}{2}(\sigma-1)(q_{xx}-r_{xx})q_x+5[(\sigma+3)r-(\sigma-1)q]r_{xxx}+5(\sigma-1)(q-r)q_{xxx}=0.\label{symKKb}
\end{align}

We use the symmetrical versions of systems to obtain nonlocal reductions of them which will be the subject of the next section.

\section{Nonlocal reductions}

In the last decade there has been intensive interest to obtain new integrable nonlocal equations \cite{abl1}-\cite{gur3}. Here we give some new nonlocal equations of fifth order, namely nonlocal SK and nonlocal KK equations.
The above symmetrical versions of SK (\ref{symSKa}), (\ref{symSKb}), and KK (\ref{symKKa}), (\ref{symKKb}) systems are good candidates to obtain new nonlocal integrable equations of fifth order.

\vspace{0.3cm}
\noindent
\noindent \textbf{(1)} Nonlocal SK equations:

Consider the symmetrical SK system (\ref{symSKa}) and (\ref{symSKb}).

\noindent (a) $r(x,t)=q(\varepsilon_1x,\varepsilon_2t)$, $\varepsilon_1^2=\varepsilon_2^2=1$.

When we apply this real nonlocal reduction on the SK system (\ref{symSKa}) and (\ref{symSKb}) we get the condition $\varepsilon_1\varepsilon_2=1$ for consistency.
Therefore here we have only one nonlocal reduction $r=q(-x,-t)$ which reduces the system (\ref{symSKa}), (\ref{symSKb}) to the following nonlocal
space-time reversal SK equation
\begin{align}
aq_t+&q_{5x}+\frac{5}{2}[(\sigma+3)q_{xx}-(\sigma-1)q_{xx}^{\varepsilon}]q_x+10[(\sigma+1)q^2-(\sigma-1)qq^{\varepsilon}]q_x
\nonumber\\&+5(\sigma-1)((q^{\varepsilon})^2-q^2)q_x^{\varepsilon}+\frac{5}{2}(\sigma-1)(q_{xx}^{\varepsilon}-q_{xx})q_x^{\varepsilon}
\nonumber\\&+\frac{5}{2}[(\sigma+3)q-(\sigma-1)q^{\varepsilon}]q_{xxx}
+\frac{5}{2}(\sigma-1)(q^{\varepsilon}-q)q_{xxx}^{\varepsilon}=0,
\end{align}
where $q^{\varepsilon}=q(-x,-t)$.

\noindent (b) $r(x,t)=\bar{q}(\varepsilon_1x,\varepsilon_2t)$, $\varepsilon_1^2=\varepsilon_2^2=1$.

Applying the complex nonlocal reduction $r(x,t)=\bar{q}(\varepsilon_1x,\varepsilon_2t)$ to the symmetrical SK system (\ref{symSKa}) and (\ref{symSKb}) yields the constraint
\begin{equation}
\bar{a}\varepsilon_1\varepsilon_2=a
\end{equation}
for consistency. The system (\ref{symSKa}), (\ref{symSKb}) reduces to nonlocal space reversal SK equation for $(\varepsilon_1,\varepsilon_2)=(-1,1)$ with $a=-\bar{a}$; nonlocal time reversal SK equation for $(\varepsilon_1,\varepsilon_2)=(1,-1)$ with $a=-\bar{a}$; nonlocal space-time reversal equation SK for $(\varepsilon_1,\varepsilon_2)=(-1,-1)$ with $a=\bar{a}$
given by
\begin{align}
aq_t+&q_{5x}+\frac{5}{2}[(\sigma+3)q_{xx}-(\sigma-1)\bar{q}_{xx}^{\varepsilon}]q_x+10[(\sigma+1)q^2-(\sigma-1)q\bar{q}^{\varepsilon}]q_x
\nonumber\\&+5(\sigma-1)((\bar{q}^{\varepsilon})^2-q^2)\bar{q}_x^{\varepsilon}+\frac{5}{2}(\sigma-1)(\bar{q}_{xx}^{\varepsilon}-q_{xx})\bar{q}_x^{\varepsilon}
\nonumber\\&+\frac{5}{2}[(\sigma+3)q-(\sigma-1)\bar{q}^{\varepsilon}]q_{xxx}
+\frac{5}{2}(\sigma-1)(\bar{q}^{\varepsilon}-q)\bar{q}_{xxx}^{\varepsilon}=0,
\end{align}
where $\bar{q}^{\varepsilon}=\bar{q}(\varepsilon_1x,\varepsilon_2t)$, $\varepsilon_1^2=\varepsilon_2^2=1$. Hence the above equation consists of three different reductions representing different nonlocal complex SK equations.
\vspace{0.3cm}
\noindent

\noindent \textbf{(2)} Nonlocal KK equations:

Consider the symmetrical KK system (\ref{KKa}) and (\ref{KKb}).

\noindent (a) $r(x,t)=q(\varepsilon_1x,\varepsilon_2t)$, $\varepsilon_1^2=\varepsilon_2^2=1$.

Similar to the symmetrical SK system, applying the reduction $r(x,t)=q(\varepsilon_1x,\varepsilon_2t)$ to the symmetrical KK system (\ref{KKa}) and (\ref{KKb})
gives the constraint $\varepsilon_1\varepsilon_2=1$ for consistency. Hence we have only one real nonlocal reduction $r=q(-x,-t)$ which reduces the system (\ref{symKKa}) and (\ref{symKKb}) to the nonlocal
space-time reversal KK equation
\begin{align}
aq_t+&q_{5x}+\frac{25}{2}[(\sigma+3)q_{xx}-(\sigma-1)q_{xx}^{\varepsilon}]q_x+40[(\sigma+1)q^2-(\sigma-1)qq^{\varepsilon}]q_x
\nonumber\\
&+20(\sigma-1)((q^{\varepsilon})^2-q^2)q_x^{\varepsilon}+\frac{25}{2}(\sigma-1)(q_{xx}^{\varepsilon}-q_{xx})q_x^{\varepsilon}\nonumber\\
&+5[(\sigma+3)q-(\sigma-1)q^{\varepsilon}]q_{xxx}
+5(\sigma-1)(q^{\varepsilon}-q)q_{xxx}^{\varepsilon}=0,
\end{align}
where $q^{\varepsilon}=q(-x,-t)$.

\noindent (b) $r(x,t)=\bar{q}(\varepsilon_1x,\varepsilon_2t)$, $\varepsilon_1^2=\varepsilon_2^2=1$.

When we apply this complex nonlocal reduction to the symmetrical KK system (\ref{symKKa}) and (\ref{symKKb}) we obtain the condition $\bar{a}\varepsilon_1\varepsilon_2=a$
for consistency as in the SK system case. The KK system (\ref{symKKa}), (\ref{symKKb}) reduces to nonlocal space reversal KK equation for $(\varepsilon_1,\varepsilon_2)=(-1,1)$ with $a=-\bar{a}$; nonlocal time reversal KK equation for $(\varepsilon_1,\varepsilon_2)=(1,-1)$ with $a=-\bar{a}$; nonlocal space-time reversal KK equation for $(\varepsilon_1,\varepsilon_2)=(-1,-1)$ with $a=\bar{a}$. Explicitly, we have
\begin{align}
aq_t+&q_{5x}+\frac{25}{2}[(\sigma+3)q_{xx}-(\sigma-1)\bar{q}_{xx}^{\varepsilon}]q_x+40[(\sigma+1)q^2-(\sigma-1)q\bar{q}^{\varepsilon}]q_x
\nonumber\\
&+20(\sigma-1)((\bar{q}^{\varepsilon})^2-q^2)\bar{q}_x^{\varepsilon}+\frac{25}{2}(\sigma-1)(\bar{q}_{xx}^{\varepsilon}-q_{xx})\bar{q}_x^{\varepsilon}\nonumber\\
&+5[(\sigma+3)q-(\sigma-1)\bar{q}^{\varepsilon}]q_{xxx}
+5(\sigma-1)(\bar{q}^{\varepsilon}-q)\bar{q}_{xxx}^{\varepsilon}=0,
\end{align}
where $\bar{q}^{\varepsilon}=\bar{q}(\varepsilon_1x,\varepsilon_2t)$, $\varepsilon_1^2=\varepsilon_2^2=1$. Then the above equation consists of three different reductions representing different nonlocal complex KK equations.

By using the reduction formulas  (1).a, (1).b, (2).a, and (2).b, and the recursion operators of SK and KK systems we can obtain the recursions operators (\ref{SK-rec}) and (\ref{KK-rec}) of the nonlocal SK and nonlocal KK equations, respectively.

\section{Shifted nonlocal reductions}

After quite a few works on integrable nonlocal reductions, Ablowitz and Musslimani generalized standard nonlocal reductions to shifted nonlocal reductions in \cite{AbMu4} as
\begin{equation}
r=q(\varepsilon_1x+x_0,\varepsilon_2t+t_0),\, x_0,\, t_0\in \mathbb{C}, \quad r=\bar{q}(\varepsilon_1x+x_0,\varepsilon_2t+t_0),\, x_0,\, t_0\in \mathbb{R},
\end{equation}
for $\varepsilon_1^2=\varepsilon_2^2=1$. It is obvious that if $x_0=t_0=0$, the shifted reductions become standard (unshifted) nonlocal reductions. There are also several works on integrable shifted nonlocal equations and their different type of solutions obtained by various type of methods \cite{AbMu5}-\cite{WT}.

Here we give shifted nonlocal SK and KK equations by applying the above shifted nonlocal reductions to the symmetrical SK system (\ref{symSKa}), (\ref{symSKb}), and the
symmetrical KK system (\ref{symKKa}), (\ref{symKKb}).

\vspace{0.3cm}
\noindent
\noindent \textbf{(1)} Shifted nonlocal SK equations:

\noindent (a) $r(x,t)=q(\varepsilon_1x+x_0,\varepsilon_2t+t_0)$, $x_0, t_0 \in \mathbb{C}$, $\varepsilon_1^2=\varepsilon_2^2=1$.

Using this real shifted nonlocal reduction on the SK system (\ref{symSKa}), (\ref{symSKb}) requires the condition $\varepsilon_1\varepsilon_2=1$ to be satisfied for consistency.
Hence we have only one shifted nonlocal reduction $r=q(-x+x_0,-t+t_0)$ and the corresponding shifted nonlocal
space-time reversal SK equation is
\begin{align}
aq_t+&q_{5x}+\frac{5}{2}[(\sigma+3)q_{xx}-(\sigma-1)q_{xx}^{\varepsilon}]q_x+10[(\sigma+1)q^2-(\sigma-1)qq^{\varepsilon}]q_x
\nonumber\\&+5(\sigma-1)((q^{\varepsilon})^2-q^2)q_x^{\varepsilon}+\frac{5}{2}(\sigma-1)(q_{xx}^{\varepsilon}-q_{xx})q_x^{\varepsilon}
\nonumber\\&+\frac{5}{2}[(\sigma+3)q-(\sigma-1)q^{\varepsilon}]q_{xxx}
+\frac{5}{2}(\sigma-1)(q^{\varepsilon}-q)q_{xxx}^{\varepsilon}=0,
\end{align}
where $q^{\varepsilon}=q(-x+x_0,-t+t_0)$, $x_0, t_0 \in \mathbb{C}$.

\noindent (b) $r(x,t)=\bar{q}(\varepsilon_1x+x_0,\varepsilon_2t+t_0)$, $\varepsilon_1^2=\varepsilon_2^2=1$.

Applying the complex shifted nonlocal reduction to the system (\ref{symSKa}), (\ref{symSKb}) gives the following constraint
\begin{equation}
\bar{a}\varepsilon_1\varepsilon_2=a
\end{equation}
for consistency. Therefore the symmetrical SK system reduces to shifted nonlocal SK equations represented by
\begin{align}
aq_t+&q_{5x}+\frac{5}{2}[(\sigma+3)q_{xx}-(\sigma-1)\bar{q}_{xx}^{\varepsilon}]q_x+10[(\sigma+1)q^2-(\sigma-1)q\bar{q}^{\varepsilon}]q_x
\nonumber\\&+5(\sigma-1)((\bar{q}^{\varepsilon})^2-q^2)\bar{q}_x^{\varepsilon}+\frac{5}{2}(\sigma-1)(\bar{q}_{xx}^{\varepsilon}-q_{xx})\bar{q}_x^{\varepsilon}
\nonumber\\&+\frac{5}{2}[(\sigma+3)q-(\sigma-1)\bar{q}^{\varepsilon}]q_{xxx}
+\frac{5}{2}(\sigma-1)(\bar{q}^{\varepsilon}-q)\bar{q}_{xxx}^{\varepsilon}=0,
\end{align}
where $\bar{q}^{\varepsilon}=\bar{q}(\varepsilon_1x+x_0,\varepsilon_2t+t_0)$, $x_0, t_0\in \mathbb{R}$, $\varepsilon_1^2=\varepsilon_2^2=1$. The above equation consists three shifted nonlocal SK equations; shifted nonlocal space reversal SK equation for $(\varepsilon_1,\varepsilon_2)=(-1,1)$, $a=-\bar{a}$, shifted nonlocal time reversal SK equation for $(\varepsilon_1,\varepsilon_2)=(1,-1)$, $a=-\bar{a}$, and shifted nonlocal space-time reversal equation SK equation for $(\varepsilon_1,\varepsilon_2)=(-1,-1)$, $a=\bar{a}$.

\vspace{0.3cm}

\noindent \textbf{(2)} Shifted nonlocal KK equations:

\noindent (a) $r(x,t)=q(\varepsilon_1x+x_0,\varepsilon_2t+t_0)$, $x_0, t_0 \in \mathbb{C}$, $\varepsilon_1^2=\varepsilon_2^2=1$.

Applying the reduction $r(x,t)=q(\varepsilon_1x+x_0,\varepsilon_2t+t_0)$ to the system (\ref{KKa}) and (\ref{KKb})
yields the condition $\varepsilon_1\varepsilon_2=1$ for consistency. Similar to the SK system case, we have only one real shifted nonlocal reduction $r=q(-x+x_0,-t+t_0)$ reducing the KK system (\ref{symKKa}) and (\ref{symKKb}) to the shifted nonlocal
space-time reversal KK equation
\begin{align}
aq_t+&q_{5x}+\frac{25}{2}[(\sigma+3)q_{xx}-(\sigma-1)q_{xx}^{\varepsilon}]q_x+40[(\sigma+1)q^2-(\sigma-1)qq^{\varepsilon}]q_x
\nonumber\\
&+20(\sigma-1)((q^{\varepsilon})^2-q^2)q_x^{\varepsilon}+\frac{25}{2}(\sigma-1)(q_{xx}^{\varepsilon}-q_{xx})q_x^{\varepsilon}\nonumber\\
&+5[(\sigma+3)q-(\sigma-1)q^{\varepsilon}]q_{xxx}
+5(\sigma-1)(q^{\varepsilon}-q)q_{xxx}^{\varepsilon}=0,
\end{align}
where $q^{\varepsilon}=q(-x+x_0,-t+t_0)$, $x_0, t_0 \in \mathbb{C}$.

\noindent (b) $r(x,t)=\bar{q}(\varepsilon_1x+x_0,\varepsilon_2t+t_0)$, $x_0, t_0 \in \mathbb{R}$, $\varepsilon_1^2=\varepsilon_2^2=1$.

Under the complex shifted nonlocal reduction we get the condition $\bar{a}\varepsilon_1\varepsilon_2=a$ for consistency and the symmetrical KK system (\ref{symKKa}), (\ref{symKKb}) reduces to three different shifted nonlocal KK equations given by
\begin{align}
aq_t+&q_{5x}+\frac{25}{2}[(\sigma+3)q_{xx}-(\sigma-1)\bar{q}_{xx}^{\varepsilon}]q_x+40[(\sigma+1)q^2-(\sigma-1)q\bar{q}^{\varepsilon}]q_x
\nonumber\\
&+20(\sigma-1)((\bar{q}^{\varepsilon})^2-q^2)\bar{q}_x^{\varepsilon}+\frac{25}{2}(\sigma-1)(\bar{q}_{xx}^{\varepsilon}-q_{xx})\bar{q}_x^{\varepsilon}\nonumber\\
&+5[(\sigma+3)q-(\sigma-1)\bar{q}^{\varepsilon}]q_{xxx}
+5(\sigma-1)(\bar{q}^{\varepsilon}-q)\bar{q}_{xxx}^{\varepsilon}=0,
\end{align}
where $\bar{q}^{\varepsilon}=\bar{q}(\varepsilon_1x+x_0,\varepsilon_2t+t_0)$, $x_0, t_0 \in \mathbb{R}$, $\varepsilon_1^2=\varepsilon_2^2=1$. Indeed we have shifted nonlocal space reversal KK equation for $(\varepsilon_1,\varepsilon_2)=(-1,1)$, $a=-\bar{a}$, shifted nonlocal time reversal KK equation for $(\varepsilon_1,\varepsilon_2)=(1,-1)$, $a=-\bar{a}$, and shifted nonlocal space-time reversal KK equation for $(\varepsilon_1,\varepsilon_2)=(-1,-1)$, $a=\bar{a}$.

\section{Hirota bilinearization and one-soliton solution}

It is also possible to write the Hirota bilinear forms of the extended equations. Starting with the Hirota bilinear forms of the scalar SK and KK equations and writing them for extended variable $U$ we obtain the corresponding Hirota forms of SK and KK systems.

\noindent \textbf{(i)} For the SK system (\ref{SKa}) and (\ref{SKb}).

Let $u=6( \ln(f))_{xx}$ then the Hirota bilinear form of SK equation is given as \cite{JM}, \cite{hos}, \cite{Kumar}
\begin{equation}
D_{x}\,\left(D_{t}+D_{x}^5 \right)\, \{f \cdot f\}=0.
\end{equation}
For the system of equations we make use of the above bilinearization. We let $U=6\, \left(F^{-1}\, F_{x} \right)_{x}$ where $F=f_{1}I+f_{2} \Sigma$ and $F^{-1}=\frac{1}{f_{1}^2-\sigma f_{2}^2}\, (f_{1}I-f_{2} \Sigma)$. Here $f_{1}$ and $f_{2}$ are the functions to be determined by the Hirota method. Then we get
\begin{equation}
FF_{6x}-6F_xF_{5x}+15F_{xx}F_{4x}-10F_{xxx}^2+FF_{xt}-F_xF_t=0
\end{equation}
or equivalently
\begin{equation}
D_{x}\,\left(D_{t}+D_{x}^5 \right)\, \{F \cdot F\}=0.
\end{equation}
When we write the above equation in componentwise we obtain
\begin{eqnarray}
&&D_{x}\,\left(D_{t}+D_{x}^5 \right)\{f_{1} \cdot f_{1}+\sigma f_{2} \cdot f_{2}\}=0,\label{SKhira}\\
&&D_{x}\,\left(D_{t}+D_{x}^5 \right)\{f_{1} \cdot f_{2}\}=0.\label{SKhirb}
\end{eqnarray}
Hence solving $f_{1}$ and $f_{2}$ from the above expressions and using them in $U=6\, \left(F^{-1}\, F_{x} \right)_{x}=uI+v\Sigma$, we obtain $u$ and $v$ in terms of $f_{1}$ and $f_{2}$. To find one-soliton solution of the system (\ref{SKa}), (\ref{SKb}), take $f_1=\alpha_1+\alpha_2e^{k_1x+\omega_1t+\delta_1}$ and $f_2=\alpha_3+
\alpha_4e^{k_2x+\omega_2t+\delta_2}$ for some constants $k_j, \omega_j, \delta_j$, $j=1, 2$ and $\alpha_i$, $i=1,2,3,4$. Inserting this choice into the Hirota bilinear form (\ref{SKhira}) and (\ref{SKhirb}) yields
\begin{equation}\label{omega_1}
k_2=k_1, \quad \omega_2=\omega_1=-k_1^5
\end{equation}
and
\begin{equation}\label{SK-sol}
u(x,t)=\frac{U_1(x,t)}{U_2(x,t)},\quad v(x,t)=\frac{U_3(x,t)}{U_2(x,t)}
\end{equation}
where{\small
\begin{align}\displaystyle
&U_1(x,t)=-6k_1^2\Big[(\sigma\alpha_3^2-\alpha_1^2)(\alpha_1\alpha_2e^{\delta_1}-\sigma\alpha_3\alpha_4e^{\delta_2})e^{\phi}+2(\sigma\alpha_3^2-\alpha_1^2)
(\alpha_2^2e^{2\delta_1}-\sigma\alpha_4^2e^{2\delta_2})e^{2\phi}\nonumber\\
&+(\alpha_1\alpha_2e^{\delta_1}-\sigma\alpha_3\alpha_4e^{\delta_2})
(\sigma\alpha_4^2e^{2\delta_2}-\alpha_2^2e^{2\delta_1})e^{3\phi}\Big],\\
&U_2(x,t)=\Big[(\alpha_1^2-\sigma\alpha_3^2)+2(\alpha_1\alpha_2e^{\delta_1}-\sigma\alpha_3\alpha_4e^{\delta_2})e^{\phi}
+(\alpha_2^2e^{2\delta_1}-\sigma\alpha_4^2e^{2\delta_2})e^{2\phi} \Big]^2, \\
&U_3(x,t)=6k_1^2\Big[(\sigma\alpha_3^2-\alpha_1^2)(\alpha_2\alpha_3e^{\delta_1}-\alpha_1\alpha_4e^{\delta_2})e^{\phi}
+(\alpha_2\alpha_3e^{\delta_1}-\alpha_1\alpha_4e^{\delta_1})(\alpha_2^2e^{2\delta_1}-\sigma\alpha_4^2e^{2\delta_2})e^{3\phi}\Big],
\end{align}}
where $\phi=k_1x-k_1^5t$.

\vspace{0.3cm}

\noindent \textbf{Example 1.} Take particular values for the parameters of the solution (\ref{SK-sol}) as $k_1=\frac{1}{2},\sigma=-1$, $\alpha_1=1, \alpha_2=-1, \alpha_3=2, \alpha_4=-\frac{3}{2}$, $\delta_1=\delta_2=0$. Then the solution $(u(x,t), v(x,t))$ becomes
\begin{align}
&u(x,t)=\frac{12e^{\frac{1}{2}x-\frac{1}{32}t}[65e^{\frac{1}{2}x-\frac{1}{32}t}-26e^{x-\frac{1}{16}t}-40]}
{[32e^{\frac{1}{2}x-\frac{1}{32}t}-13e^{x-\frac{1}{16}t}-20]^2},
&v(x,t)=\frac{-3e^{\frac{1}{2}x-\frac{1}{32}t}[13e^{x-\frac{1}{16}t}-20]}
{[32e^{\frac{1}{2}x-\frac{1}{32}t}-13e^{x-\frac{1}{16}t}-20]^2}.
\end{align}
The graphs of the above solutions are given in Figure 1.
\begin{center}
\begin{figure}[h!]
    \centering
    \subfigure[]{\includegraphics[width=0.30\textwidth]{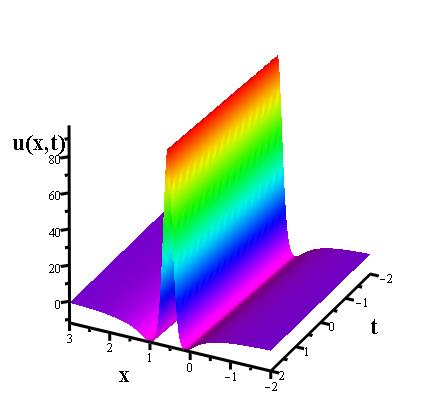}}\hspace{2.5cm}
    \subfigure[]{\includegraphics[width=0.30\textwidth]{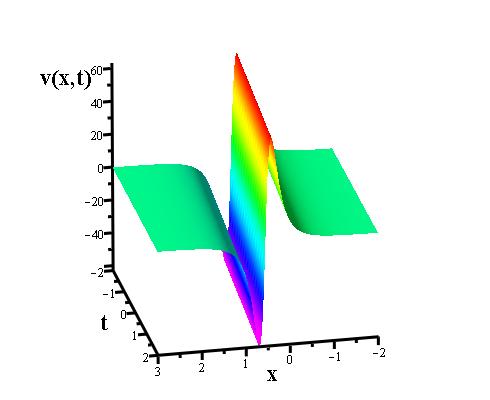}}
        \caption{One-soliton solution $(u(x,t),v(x,t))$ of the SK system (\ref{SKa}), (\ref{SKb}) for $k_1=\frac{1}{2}$, $\sigma=-1$, $\alpha_1=1, \alpha_2=-1, \alpha_3=2, \alpha_4=-\frac{3}{2}$, $\delta_1=\delta_2=0$.}
    \end{figure}
\end{center}
\squeezeup

\noindent \textbf{(ii)} For the KK system (\ref{KKa}) and (\ref{KKb}).

Letting $u=\frac{3}{2}( \ln(f))_{xx}$ yields the Hirota bilinear form of KK equation as \cite{JM}, \cite{Parker1}, \cite{Parker2}
\begin{align}
&(16D_xD_t+D_x^6)\{f\cdot f\}+20D_x^2\{f\cdot g \}=0,\\
&D_x^4 \{f\cdot f\}-\frac{4}{3}fg=0,
\end{align}
where $g$ is an auxiliary function. Similar to the SK system, we let $U=\frac{3}{2}\, \left(F^{-1}\, F_{x} \right)_{x}$ where $F=f_{1}I+f_{2} \Sigma$. We determine $f_{1}$ and $f_{2}$ by the Hirota method. We get
\begin{align}
&32(FF_{xt}-F_xF_t)+2(FF_{6x}-6F_xF_{5x}+15F_{xx}F_{4x}-10F_{xxx}^2)\nonumber\\
&\hspace{7cm}+20(F_{xx}G-2F_xG_x+FG_{xx})=0,\\
&2(FF_{4x}-4F_xF_{xxx}+6F_{xx}^2)-\frac{4}{3}FG=0,
\end{align}
which is equivalent to
\begin{align}
&(16D_xD_t+D_x^6)\{F\cdot F\}+20D_x^2\{F\cdot G \}=0,\\
&D_x^4 \{F\cdot F\}-\frac{4}{3}FG=0,
\end{align}
where $G=g_{1} I+g_{2} \Sigma$. In componentwise we have
\begin{align}
&(16D_xD_t+D_x^6)\{f_1\cdot f_1+\sigma f_2\cdot f_2\}+20D_x^2\{f_1\cdot g_1+\sigma f_2\cdot g_2\}=0,\label{KKhira}\\
&(16D_xD_t+D_x^6)\{f_1\cdot f_2+ f_2\cdot f_1\}+20D_x^2\{f_2\cdot g_1+ f_1\cdot g_2\}=0,\label{KKhirb}\\
&D_x^4\{f_1\cdot f_1+\sigma f_2\cdot f_2\}-\frac{4}{3}(f_1g_1+\sigma f_2 g_2)=0,\label{KKhirc}\\
&D_x^4\{f_2\cdot f_1+ f_1\cdot f_2\}-\frac{4}{3}(f_2g_1+f_1 g_2)=0.\label{KKhird}
\end{align}
To obtain one-soliton solution of the system (\ref{KKa}), (\ref{KKb}) we take $f_1=\alpha_0+\alpha_1e^{\theta_1}+\alpha_2e^{2\theta_1}$, $f_2=\alpha_3+\alpha_4e^{\theta_1}$, $g_1=\alpha_5e^{\theta_1}$, and $g_2=\alpha_6e^{\theta_1}$, where $\theta_1=k_1x+\omega_1 t+\delta_1$ for some constants $k_1, \omega_1, \delta_1$, and $\alpha_j$, $j=1,\ldots, 6$. From the above system we get $\omega_1=-k_1^5$ and the following constraints:
\begin{equation}\displaystyle
\alpha_0=\frac{\alpha_3(\sigma\alpha_4^2+\alpha_1^2)}{2\alpha_1\alpha_4},\,\, \alpha_2=\frac{\alpha_1\alpha_4}{8\alpha_3},\,\, \alpha_5=\frac{3}{2}\alpha_1k_1^4,\,\, \alpha_6=\frac{3}{2}\alpha_4k_1^4.
\end{equation}
Hence one-soliton solution of the KK system (\ref{KKa}), (\ref{KKb}) is given by the pair $(u,v)$
\begin{equation}\displaystyle
u=\frac{W_1}{W_2},\quad v=\frac{W_3}{W_2}, \label{KK-sol}
\end{equation}
where{\small
\begin{align}\displaystyle
&W_1=-12k_1^2\alpha_1^2\alpha_3\alpha_4 \Big[32\alpha_3^5\alpha_4(7\sigma\alpha_4^2-9\alpha_1^2)(\sigma\alpha_4^2-\alpha_1^2)^2e^{\theta_1}-16\alpha_1^2\alpha_3^4\alpha_4^2
(\sigma\alpha_4^2-\alpha_1^2)(23\sigma\alpha_4^2-27\alpha_1^2)e^{2\theta_1}\nonumber\\
&
-16\alpha_1^2\alpha_3^3\alpha_4^3(\sigma\alpha_4^2-\alpha_1^2)(\sigma\alpha_4^2-17\alpha_1^2)e^{3\theta_1}
+4\alpha_1^4\alpha_3^2\alpha_4^4(23\sigma\alpha_4^2-27\alpha_1^2)e^{4\theta_1}
\nonumber\\
&+2\alpha_1^4\alpha_3\alpha_4^5(7\sigma\alpha_4^2-9\alpha_1^2)e^{5\theta_1}-\alpha_4^6\alpha_1^6e^{6\theta_1}+64\alpha_3^6(\sigma\alpha_4^2-\alpha_1^2)^3\Big],\\
&W_2=\Big[64\alpha_1^2\alpha_3^3\alpha_4(\sigma\alpha_4^2-\alpha_1^2)+8\alpha_1^2\alpha_3^2\alpha_4^2(7\sigma\alpha_4^2-9\alpha_1^2)e^{2\theta_1}
-16\alpha_1^4\alpha_4^3\alpha_3e^{3\theta_1}-\alpha_4^4\alpha_1^4e^{4\theta_1}\nonumber\\
&-16\alpha_3^4(\sigma\alpha_4^2-\alpha_1^2)^2\Big]^2,\\
&W_3=-12k_1^2\alpha_4^2\alpha_1\alpha_3e^{\theta_1}\Big[64\alpha_1^2\alpha_3^2\alpha_4(\sigma\alpha_4^2-\alpha_1^2)^2e^{\theta_1}-16\alpha_1^2\alpha_3^4\alpha_4^2
(11\sigma\alpha_4^2-7\alpha_1^2)(\sigma\alpha_4^2-\alpha_1^2)e^{2\theta_1}
\nonumber\\
&-4\alpha_1^4\alpha_3^2\alpha_4^2(11\sigma\alpha_4^2-7\alpha_1^2)e^{4\theta_1}
-4\alpha_1^6\alpha_4^5\alpha_3e^{5\theta_1}-\alpha_4^6\alpha_1^6e^{6\theta_1}-64\alpha_3^6(\sigma\alpha_4^2-\alpha_1^2)^3\Big],
\end{align}}
where $\theta_1=k_1x-k_1^5t+\delta_1$.

\vspace{0.3cm}

\noindent \textbf{Example 2.} Consider the particular values for the parameters of the solution (\ref{KK-sol}) as $k_1=1,\sigma=-1$, $\alpha_1=\alpha_3=1, \alpha_4=2, \delta_1=0$. We have the solution $(u(x,t), v(x,t))$ where
\begin{align}
u(x,&t)=24e^{x-t}\Big[59200e^{x-t}+38080e^{2x-2t}+13440e^{3x-3t}+7616e^{4x-4t}\nonumber\\&
+2368e^{5x-5t}+64e^{6x-6t}+8000\Big]\Big/[640e^{x-t}+1184e^{2x-2t}+128e^{3x-3t}+16e^{4x-4t}+400\Big]^2,\\
v(x,&t)=-48e^{x-t}\Big[3200e^{x-t}-16320e^{2x-2t}+3264e^{4x-4t}-128e^{5x-5t}-64e^{6x-6t}+8000\Big]\nonumber\\
&\Big/\Big[640e^{x-t}+1184e^{2x-2t}+128e^{3x-3t}+16e^{4x-4t}+400\Big]^2.
\end{align}
The graphs of the above solutions are given in Figure 2.
\begin{center}
\begin{figure}[h!]
    \centering
    \subfigure[]{\includegraphics[width=0.30\textwidth]{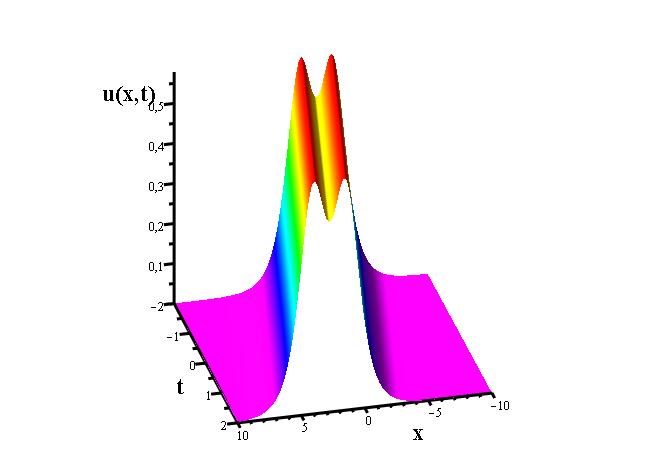}}\hspace{2.5cm}
    \subfigure[]{\includegraphics[width=0.30\textwidth]{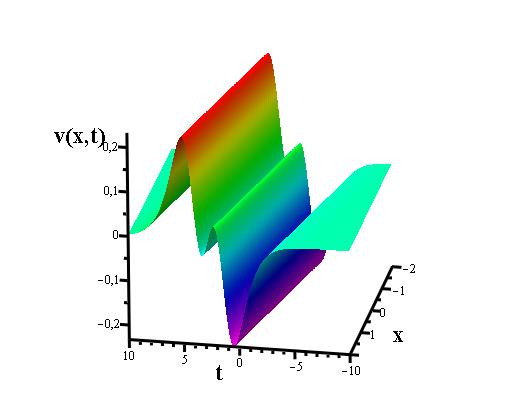}}
        \caption{One-soliton solution $(u(x,t),v(x,t))$ of the KK system (\ref{KKa}), (\ref{KKb}) for $k_1=1,\sigma=-1$, $\alpha_1=\alpha_3=1, \alpha_4=2, \delta_1=0$.}
    \end{figure}
\end{center}
\squeezeup
\noindent \textbf{Remark 2.} We note that taking different expansions for the functions $f_1$ and $f_2$ in (\ref{SKhira})
and (\ref{SKhirb}) or $f_1, f_2, g_1, g_2$ in (\ref{KKhira})-(\ref{KKhird}) may put strong conditions on the parameters
yielding $u=0$ or $v=0$. For instance, if we use $f_1=f_0+e^{k_1x+\omega_1t+\delta_1}$, $f_2=g_0$, where $f_0$, $g_0$ are constants, in
the Hirota bilinear form (\ref{SKhira}), (\ref{SKhirb}) we get $\omega_1=-k_1^5$ and also $g_0=0$ making $v=0$. In this case the SK system (\ref{SKa}), (\ref{SKb})
for the dynamical variables $u$ and $v$ reduces to the SK equation (\ref{SKeqn}) for $u$.

Since the expressions are quite longer we shall not display two- and three-soliton solutions of the SK and KK systems here. The more interesting case is the soliton solutions of the nonlocal SK and KK equations presented in Section 3. They can be easily obtained by using the above soliton solutions of the systems with the reduction formulas (1).a, (1).b, (2).a, and (2).b presented in Section 3. These equations will restrict parameters in the soliton solutions (\ref{SK-sol}) and (\ref{KK-sol}).
We shall discuss two- and three-soliton solutions of the SK and KK systems and soliton solutions of the nonlocal SK and KK equations in a forthcoming publication.

\section{Concluding remarks}
In this work we obtained systems of integrable equations with their recursion operators from scalar integrable equations by applying ${\cal M}_{2}$-extension.  We used this method for SK  and KK equations and obtained SK system and KK system of equations, respectively. Applying the nonlocal reductions to the symmetric versions of SK and KK systems we obtained eight different new standard nonlocal and eight different new shifted nonlocal integrable differential equations. We presented also one-soliton solutions of the SK and KK systems.

\section{Acknowledgment}
  This work is partially supported by the Scientific
and Technological Research Council of Turkey (T\"{U}B\.{I}TAK).\\

\end{document}